\documentclass[a4paper,11pt]{article}
\usepackage{pos}

\usepackage{subfigure}
\usepackage{tikz}
\usetikzlibrary{positioning,shapes,shadows,arrows,decorations,decorations.pathmorphing,calc}

\newcommand{\dd}{\mathrm{d}}
\newcommand{\LL}{\mathcal{L}}
\newcommand{\ord}{\mathcal{O}}
\newcommand{\fig}{figure\ }

\title{Investigation of the hadronic light-by-light contribution to the muon $g-2$ using staggered fermions}
\ShortTitle{HLbL contribution to the muon $g-2$}

\author*[a]{Christian Zimmermann}
\author[a]{Antoine Gérardin}

\affiliation[a]{Aix Marseille Univ, Université de Toulon, CNRS, CPT, Marseille, France}

\emailAdd{christian.zimmermann@univ-amu.fr}
\emailAdd{antoine.gerardin@univ-amu.fr}

\abstract{Hadronic contributions dominate the uncertainty of the standard model prediction for the anomalous magnetic moment of the muon. In this work, we describe an ongoing lattice calculation of the hadronic light-by-light contribution, performed with staggered fermions. The presence of quarks with different tastes complicates the analysis of the position-space correlation function. We present a suitable adaption of the "Mainz method". As a first numerical test, we reproduce the well-known lepton-loop contribution. Results at a single lattice spacing for the light quark contribution, using two volumes, are then discussed. Our study of the long distance behavior and finite-volume effects is supplemented by considering the contribution of the light pseudoscalar-pole. The corresponding transition form factors have been evaluated in previous simulations on the same ensembles.}

\FullConference{The 40th International Symposium on Lattice Field Theory (Lattice 2023)\\
July 31st - August 4th, 2023\\
Fermi National Accelerator Laboratory\\}

\begin{document}
\maketitle

\section{Introduction}

Tensions between measurements in experiments and results of theoretical calculations of the anomalous magnetic moment can give hints for physics beyond the standard model, which is why the research on that topic remains a field of large interest. The anomalous magnetic moment has been determined precisely in experiments by BNL \cite{Muong-2:2006rrc} and Fermilab \cite{Muong-2:2023cdq}, which present a tension with standard model estimates. Recent theoretical results based on the dispersive approach and lattice calculations differ from the experimental values by up to $4.2\sigma$ \cite{Aoyama:2020ynm,Borsanyi:2020mff}. On the theoretical side, there are several contributions to consider. Among the well known QED and electroweak sector, this includes hadronic contributions. The theory error is dominated by hadronic vacuum polarization (HVP) contributing at $\ord(\alpha^2)$, as well as the hadronic light-by-light scattering (HLbL) at $\ord(\alpha^3)$. Both have to be treated non-perturbatively. 

The present study addresses the determination of the HLbL contribution on the lattice through a direct calculation of the required four-current correlator. The evaluation of this quantity is computationally challenging. However, since it is suppressed by an additional factor of $\alpha$ compared to the HVP, a precision of $10\%$ is sufficient. A corresponding evaluation of the HLbL contribution has been already performed in the past by several groups \cite{Blum:2023vlm,Chao:2021tvp}.

In these proceedings, we shall describe our methods used for the calculation of the HLbL contribution to $a_\mu$. In particular, we address the adaption of the method developed by the Mainz group \cite{Chao:2020kwq,Chao:2021tvp} in the context of staggered fermions. Moreover, we will show some first results for light- and strange-quark contributions.

\section{Hadronic light-by-light contribution on the lattice}

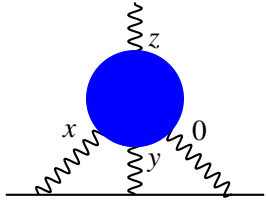
\begin{figure}
\begin{center}
\begin{tikzpicture}[scale=0.85]
\draw[thick] (-2,0) -- (2,0);
\draw[decoration={snake, segment length=5pt},decorate,thick] (-1.5,0) -- (0,1.5);
\draw[decoration={snake, segment length=5pt},decorate,thick] (0,0) -- (0,1.5);
\draw[decoration={snake, segment length=5pt},decorate,thick] (1.5,0) -- (0,1.5);
\draw[decoration={snake, segment length=5pt},decorate,thick] (0,1.5) -- (0,3);
\filldraw[blue] (0,1.5) circle (0.75);
\draw (-1,1) node {$x$};
\draw (1,1) node {$0$};
\draw (0.3,0.5) node {$y$};
\draw (0.3,2.4) node {$z$};
\end{tikzpicture}
\end{center}
\caption{Feynman diagram for the hadronic light-by-light scattering amplitude at $\ord(\alpha^3)$. The blue blob represents the hadronic four-current correlation function. Straight (wavy) lines represent the muon (photons).\label{fig:hlbl-general}}
\end{figure}

The contribution of hadronic light-by-light scattering to the anomalous magnetic moment of the muon is given by the formula \cite{Chao:2020kwq,Chao:2021tvp}:
\vspace*{-0.3cm}

\begin{align}
a_\mu^{\mathrm{HLbL}} = \frac{m_\mu e^6}{3} \int \dd x^4 \int \dd y^4 \LL_{[\rho,\sigma];\mu\nu\lambda}(x,y) \int \dd z^4 (-z_\rho) \widetilde{\Pi}_{\mu\nu\lambda\sigma}(x,y,z) \,.
\label{eq:amu-hlbl}
\end{align}
The corresponding Feynman diagram is shown in \fig\ref{fig:hlbl-general}. The electromagnetic kernel $\LL$ corresponds to the muon and the photons lines and has been determined semi-analytically in \cite{Asmussen:2022oql}. The QCD part is represented by the four-current correlator $\widetilde{\Pi}$ (blue blob), which is defined as:
\vspace*{-0.3cm}

\begin{align}
\widetilde{\Pi}(x,y,z) := \left\langle j_\mu(x) j_\nu(y) j_\lambda(0) j_\sigma(z) \right\rangle \,,
\label{eq:4c-corr}
\end{align}
with the hadronic component of the electromagnetic current $j_\mu(x)$.

\begin{figure}
\begin{center}
\begin{tikzpicture}[scale=0.4]
\draw (-3,0) node[left] {$\widetilde{\Pi} =$};
\draw[very thick] (0,2) -- (-2,0) -- (0,-2) -- (2,0) -- (0,2);
\draw[very thick,->] (-2,0) -- (-1,-1);
\draw[very thick,->] (0,2) -- (-1,1);
\draw[very thick,->] (0,-2) -- (1,-1);
\draw[very thick,->] (2,0) -- (1,1);
\draw (3,0) node {$+$};
\end{tikzpicture}
\begin{tikzpicture}[scale=0.4]
\draw[very thick] (-2,0) arc(180:90:2);
\draw[very thick] (0,2) arc(0:-90:2);
\draw[very thick,->] (-2,0) arc(180:135:2);
\draw[very thick,->] (0,2) arc(0:-45:2);
\draw[very thick] (0,-2) arc(180:90:2);
\draw[very thick] (2,0) arc(0:-90:2);
\draw[very thick,->] (0,-2) arc(180:135:2);
\draw[very thick,->] (2,0) arc(0:-45:2);
\draw (3,0) node {$+$};
\end{tikzpicture}
\begin{tikzpicture}[scale=0.4]
\draw[very thick] (0,2) -- (-2,0) -- (0,-2) -- (0,2);
\draw[very thick,->] (-2,0) -- (-1,-1);
\draw[very thick,->] (0,2) -- (-1,1);
\draw[very thick,->] (0,-2) -- (0,0);
\draw[very thick] (2,0) arc(0:-360:0.6);
\draw[very thick,->] (2,0) arc(0:-180:0.6);
\draw (3,0) node {$+$};
\end{tikzpicture}
\begin{tikzpicture}[scale=0.4]
\draw[very thick] (-2,0) arc(180:90:2);
\draw[very thick] (0,2) arc(0:-90:2);
\draw[very thick,->] (-2,0) arc(180:135:2);
\draw[very thick,->] (0,2) arc(0:-45:2);
\draw[very thick] (0,-2) arc(270:-90:0.6);
\draw[very thick,->] (0,-2) arc(270:90:0.6);
\draw[very thick] (2,0) arc(0:-360:0.6);
\draw[very thick,->] (2,0) arc(0:-180:0.6);
\draw (2.5,0) node[right] {$+$};
\end{tikzpicture}
\begin{tikzpicture}[scale=0.4]
\draw[very thick] (-2,0) arc(-180:180:0.6);
\draw[very thick,->] (-2,0) arc(180:0:0.6);
\draw[very thick] (0,2) arc(-270:90:0.6);
\draw[very thick,->] (0,2) arc(90:-90:0.6);
\draw[very thick] (0,-2) arc(270:-90:0.6);
\draw[very thick,->] (0,-2) arc(270:90:0.6);
\draw[very thick] (2,0) arc(0:-360:0.6);
\draw[very thick,->] (2,0) arc(0:-180:0.6);
\end{tikzpicture}
\end{center}
\caption{All five types of Wick contractions contributing to the four-current correlation function $\widetilde{\Pi}$. In our simulation we only consider the first two types, i.e. connected and $2+2$ diagrams.\label{fig:wick-contractions}}
\end{figure}
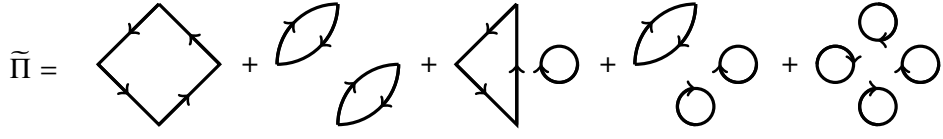

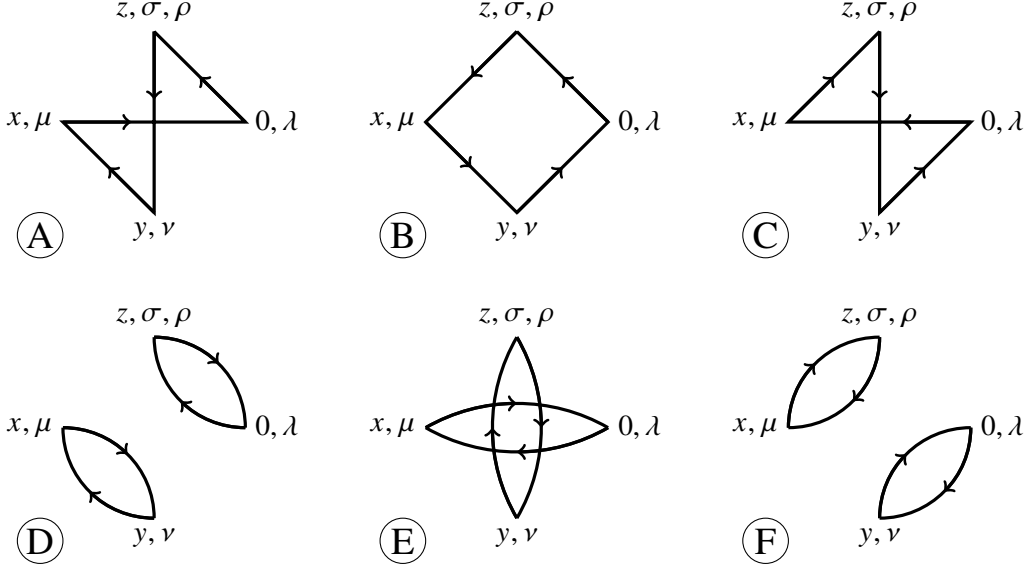
\begin{figure}
\begin{center}
\begin{tikzpicture}[scale=0.6]
\draw[very thick] (0,2) -- (0,-2) -- (-2,0) -- (2,0) -- (0,2);
\draw[very thick,->] (0,-2) -- (-1,-1);
\draw[very thick,->] (0,2) -- (0,0.5);
\draw[very thick,->] (2,0) -- (1,1);
\draw[very thick,->] (-2,0) -- (-0.5,0);
\draw (0,2) node[above] {$z,\sigma,\rho$};
\draw (-2,0) node[left] {$x,\mu$};
\draw (2,0) node[right] {$0,\lambda$};
\draw (0,-2) node[below] {$y,\nu$};
\draw (-2.5,-2.5) circle(0.5) node {\Large A};
\end{tikzpicture}
\hspace*{0.5cm}
\begin{tikzpicture}[scale=0.6]
\draw[very thick] (0,2) -- (-2,0) -- (0,-2) -- (2,0) -- (0,2);
\draw[very thick,->] (-2,0) -- (-1,-1);
\draw[very thick,->] (0,2) -- (-1,1);
\draw[very thick,->] (0,-2) -- (1,-1);
\draw[very thick,->] (2,0) -- (1,1);
\draw (0,2) node[above] {$z,\sigma,\rho$};
\draw (-2,0) node[left] {$x,\mu$};
\draw (2,0) node[right] {$0,\lambda$};
\draw (0,-2) node[below] {$y,\nu$};
\draw (-2.5,-2.5) circle(0.5) node {\Large B};
\end{tikzpicture}
\hspace*{0.5cm}
\begin{tikzpicture}[scale=0.6]
\draw[very thick] (0,2) -- (-2,0) -- (2,0) -- (0,-2) -- (0,2);
\draw[very thick,->] (-2,0) -- (-1,1);
\draw[very thick,->] (0,2) -- (0,0.5);
\draw[very thick,->] (0,-2) -- (1,-1);
\draw[very thick,->] (2,0) -- (0.5,0);
\draw (0,2) node[above] {$z,\sigma,\rho$};
\draw (-2,0) node[left] {$x,\mu$};
\draw (2,0) node[right] {$0,\lambda$};
\draw (0,-2) node[below] {$y,\nu$};
\draw (-2.5,-2.5) circle(0.5) node {\Large C};
\end{tikzpicture} \\
\vspace*{0.5cm}
\begin{tikzpicture}[scale=0.6]
\draw[very thick] (0,-2) arc(270:180:2);
\draw[very thick] (-2,0) arc(90:0:2);
\draw[very thick,->] (0,-2) arc(270:225:2);
\draw[very thick,->] (-2,0) arc(90:45:2);
\draw[very thick] (0,2) arc(90:0:2);
\draw[very thick] (2,0) arc(270:180:2);
\draw[very thick,->] (0,2) arc(90:45:2);
\draw[very thick,->] (2,0) arc(270:225:2);
\draw (0,2) node[above] {$z,\sigma,\rho$};
\draw (-2,0) node[left] {$x,\mu$};
\draw (2,0) node[right] {$0,\lambda$};
\draw (0,-2) node[below] {$y,\nu$};
\draw (-2.5,-2.5) circle(0.5) node {\Large D};
\end{tikzpicture}
\hspace*{0.5cm}
\begin{tikzpicture}[scale=0.6]
\draw[very thick] (0,-2) arc(210:150:4);
\draw[very thick] (0,2) arc(30:-30:4);
\draw[very thick] (-2,0) arc(120:60:4);
\draw[very thick] (2,0) arc(300:240:4);
\draw[very thick,->] (0,-2) arc(210:180:4);
\draw[very thick,->] (0,2) arc(30:0:4);
\draw[very thick,->] (-2,0) arc(120:90:4);
\draw[very thick,->] (2,0) arc(300:270:4);
\draw (0,2) node[above] {$z,\sigma,\rho$};
\draw (-2,0) node[left] {$x,\mu$};
\draw (2,0) node[right] {$0,\lambda$};
\draw (0,-2) node[below] {$y,\nu$};
\draw (-2.5,-2.5) circle(0.5) node {\Large E};
\end{tikzpicture}
\hspace*{0.5cm}
\begin{tikzpicture}[scale=0.6]
\draw[very thick] (-2,0) arc(180:90:2);
\draw[very thick] (0,2) arc(0:-90:2);
\draw[very thick,->] (-2,0) arc(180:135:2);
\draw[very thick,->] (0,2) arc(0:-45:2);
\draw[very thick] (0,-2) arc(180:90:2);
\draw[very thick] (2,0) arc(0:-90:2);
\draw[very thick,->] (0,-2) arc(180:135:2);
\draw[very thick,->] (2,0) arc(0:-45:2);
\draw (0,2) node[above] {$z,\sigma,\rho$};
\draw (-2,0) node[left] {$x,\mu$};
\draw (2,0) node[right] {$0,\lambda$};
\draw (0,-2) node[below] {$y,\nu$};
\draw (-2.5,-2.5) circle(0.5) node {\Large F};
\end{tikzpicture}
\end{center}
\caption{Depiction of all three independent connected Wick contractions (first line), as well as all $2+2$ Wick contractions (second line).\label{fig:leading-wick}}
\end{figure}

The four-current correlator \eqref{eq:4c-corr} decomposes into several types of Wick contractions, which are shown in \fig\ref{fig:wick-contractions}. 
In the present study, we shall consider the two leading types, i.e.\ connected diagrams and diagrams containing two-current loops only (2+2). There are six connected diagrams that appear as complex conjugate pairs. One diagram of each pair is depicted in \fig\ref{fig:leading-wick}, first line. 
The second line shows the three 2+2 diagrams. We are able to exploit translation invariance to relate each of the connected diagrams to one single connected diagram. Here it is advantageous to choose the diagram $\mathrm{A}$. The overall connected contribution can be written as:
\vspace*{-0.3cm}

\begin{align}
a_\mu^{\mathrm{HLbL,c}} &= \frac{m_\mu e^6}{3} 2\pi^2 \sum_{|y|} |y|^3 \sum_{x,z} \LL_{[\rho,\sigma];\mu\nu\lambda}(x,y)\ (x_\rho - 3 z_\rho)\ \widetilde{\Pi}^{(\mathrm{A})}_{\mu\nu\lambda\sigma}(x,y,z) \nonumber\\
&= \frac{m_\mu e^6}{3} 2\pi^2 \sum_{|y|} |y|^3 f(y) \,.
\label{eq:conn}
\end{align}
Notice that the kernel is not unique. In the present study we use a fully symmetrized version of the kernel which satisfies $\LL_{[\rho\sigma];\mu\nu\lambda}(x,y) = \LL_{[\rho\sigma];\nu\mu\lambda}(y,x) = \LL_{[\rho\sigma];\lambda\nu\mu}(-x,y-x)$ and differs from the version used in \cite{Chao:2021tvp}. In the continuum limit, the function $f$ is rotationally invariant and the sum over $y$ can be reduced to a one-dimensional integral over $|y|$.

The quark disconnected contribution can be treated similarly. In this case, we express everything in terms of the contractions $\mathrm{D}$ or $\mathrm{E}$, respectively. Our final result is given by taking the average of both versions:
\vspace*{-0.3cm}

\begin{align}
a_\mu^{\mathrm{HLbL,d1}} &= \frac{m_\mu e^6}{3} 2\pi^2 \sum_{|y|} |y|^3 \sum_{x,z} \LL_{[\rho,\sigma];\mu\nu\lambda}(x,y)\ (y_\rho - 3 z_\rho)\ \widetilde{\Pi}^{(\mathrm{E})}_{\mu\nu\lambda\sigma}(x,y,z) \,,
\nonumber \\
a_\mu^{\mathrm{HLbL,d2}} &= \frac{m_\mu e^6}{3} 2\pi^2 \sum_{|y|} |y|^3 \sum_{x,z} \LL_{[\rho,\sigma];\mu\nu\lambda}(x,y)\ (x_\rho + y_\rho - 3 z_\rho)\ \widetilde{\Pi}^{(\mathrm{D})}_{\mu\nu\lambda\sigma}(x,y,z) \,.
\label{eq:disc}
\end{align}

\section{Staggered tastes and position space}

One complication we have to deal with is the behavior of observables in position space that are calculated on the lattice using the staggered discretization. In this case, we have contributions of different tastes, which are distributed within $2^4$ sub-lattices. Thus, $f(y)$ receives contributions from 16 partners with different quantum numbers. We note that the contribution from additional partners would vanish if one would be able to sum over $y$ explicitly.

As already pointed out before, we sample only a few points for a set of absolute values $|y|$. This forces us to project onto the taste singlet for each of the chosen points. This is realized by considering:
\vspace*{-0.3cm}

\begin{align}
f^{(\mathcal{S})}(y) = \prod_\mu \mathcal{S}_\mu f(y) \,,
\end{align}
where $\mathcal{S}_\mu$ is a suitable "smearing" function \cite{Lepage:2011vr}. The smeared function $f^{(\mathcal{S})}(y)$ has a discretization error specific to the applied smearing $\mathcal{S}$:
\vspace*{-0.3cm}

\begin{align}
\mathcal{S}^{(1)}_\mu f(y) &:= \frac{f(y)+f(y+\hat{\mu})}{2} \phantom{f(y-\hat{\mu})+2}\quad 
\Rightarrow \ord(a)  \nonumber \\
\mathcal{S}^{(2)}_\mu f(y) &:= \frac{f(y-\hat{\mu})+2f(y)+f(y+\hat{\mu})}{4}\quad
\Rightarrow \ord(a^2) \,.
\end{align}
\begin{figure}
\begin{center}
\subfigure[lepton loop integrand\label{fig:free-test-integrand}]{
\includegraphics[scale=0.55, clip, trim=0 0 0 0.55cm]{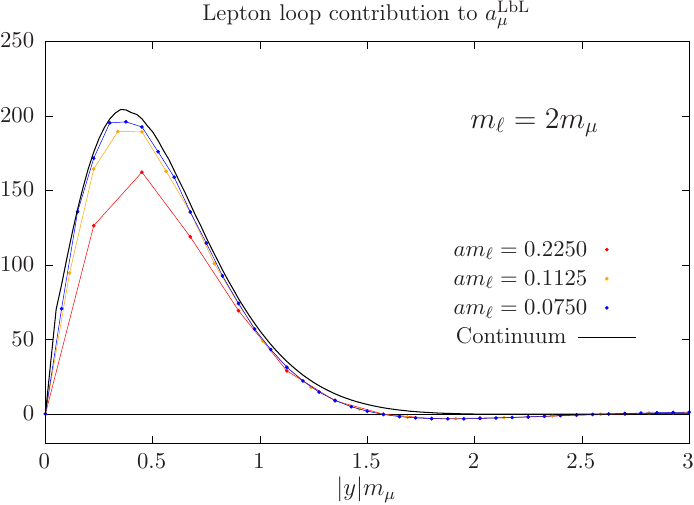}
\hspace*{0.2cm}
}
\subfigure[lepton loop continuum extrapolation\label{fig:free-test-continuum}]{
\includegraphics[scale=0.55, clip, trim=0 0 0 0.5cm]{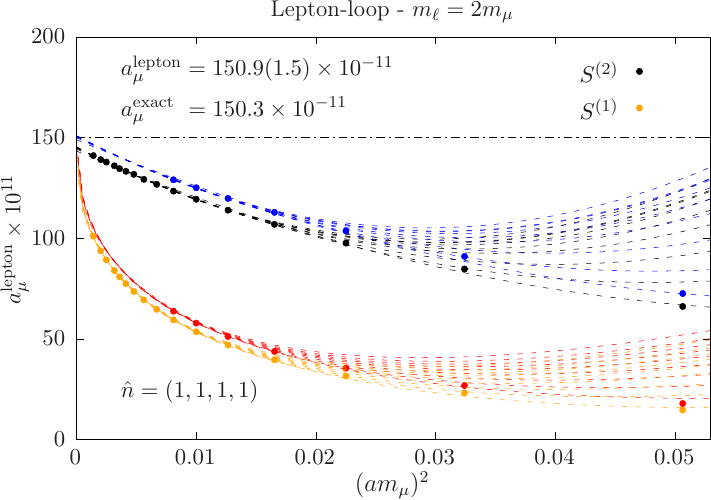}
}
\end{center}
\vspace*{-0.75cm}
\caption{Calculation of the lepton loop contribution to $a_\mu$. Left: The integrand as a function of $|y|$ for smearing $\mathcal{S}^{(2)}$. Right: Continuum extrapolation for the two smearing versions. The different lines of the same color indicate different fit ranges. The red and blue lines correspond to the larger lattices.\label{fig:free-test}}
\end{figure}
As a cross-check, we calculate the lepton loop on the lattice for a lepton mass of $m_\ell = 2m_\mu$ using the smearing function $\mathcal{S}^{(1)}$ and $\mathcal{S}^{(2)}$, respectively. In \fig\ref{fig:free-test-integrand} we show the integrand for smearing $\mathcal{S}^{(2)}$ and compare it to the analytic continuum calculation. It appears that staggered oscillations are in fact not visible. Figure \ref{fig:free-test-continuum} shows the continuum extrapolation of the lepton loop for the two different smearing versions, where we use the fit ans\"atze
\vspace*{-0.3cm}

\begin{align}
\mathcal{S}^{(1)}_\mu:\quad a^{\mathrm{lepton}}_\mu(a) &= c_0 + c_1 a + c_2 a^2 + c_4 a^4 \label{eq:fit-S1}\,, \\
\mathcal{S}^{(2)}_\mu:\quad a^{\mathrm{lepton}}_\mu(a) &= c_0 + c_2 a^2 + c_4 a^4 \label{eq:fit-S2} \,,
\end{align}
depending on the employed smearing function. The resulting curves clearly exhibit an $a$- or $a^2$-dependence at small lattice spacings for $\mathcal{S}^{(1)}$ or $\mathcal{S}^{(2)}$, respectively. For the latter case, we obtain as result $a_\mu^{\mathrm{lepton,latt}} = 150.9(1.5)\times 10^{-11}$, which is pretty close to the well known analytic value $a_\mu^{\mathrm{lepton}} = 150.3\times 10^{-11}$. The error is a naive estimate of the systematic uncertainty associated to the continuum extrapolation. It is obtained by performing cuts in the lattice spacing.

\section{Lattice setup and results}

Our simulations are performed on a set of lattices generated by the Budapest-Marseille-Wuppertal (BMW) collaboration employing a tree-level improved gauge action with staggered fermions and four iterations of stout smearing \cite{Borsanyi:2020mff}. 

\begin{figure}
\begin{center}
\subfigure[\label{fig:finvol-strange}]{
\includegraphics[scale=0.35, clip, trim=0 -1.3cm 0 3cm]{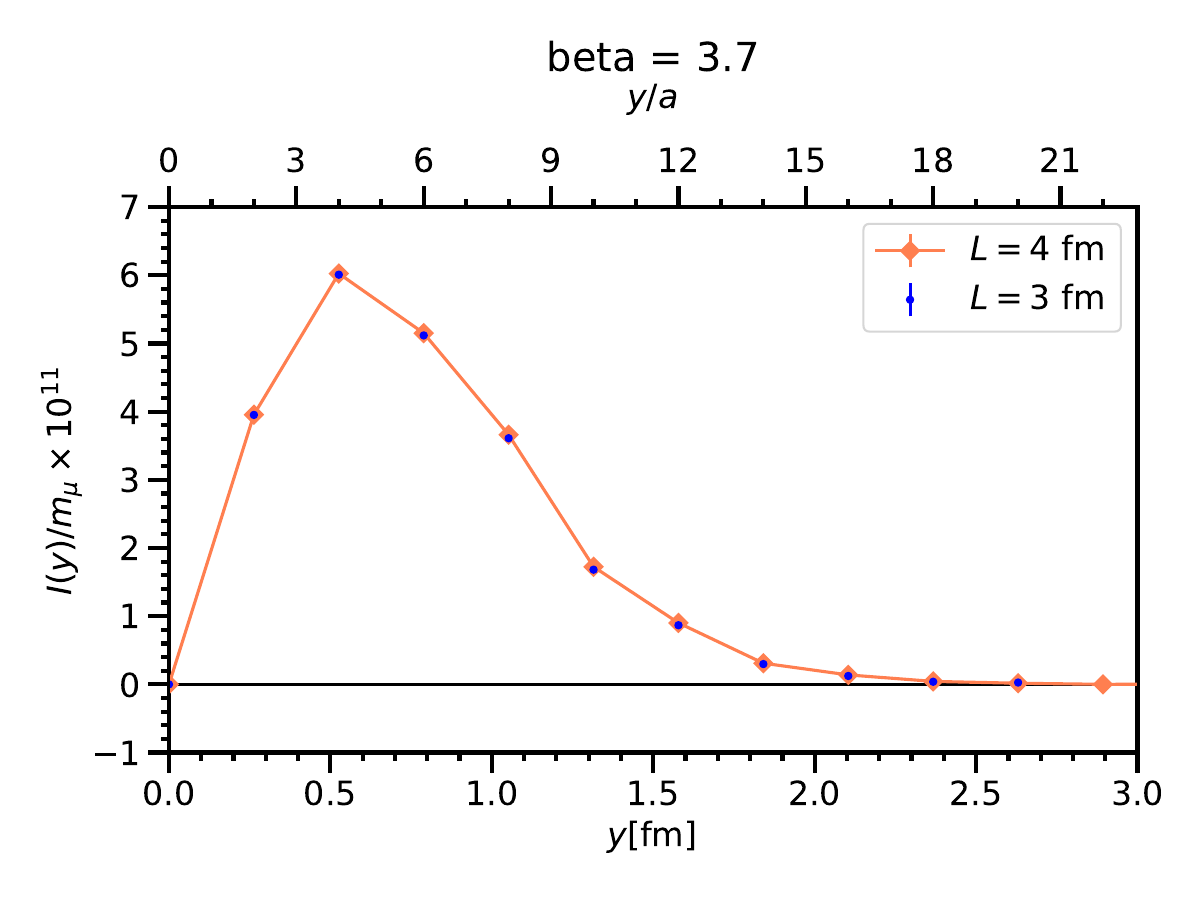}
}
\subfigure[\label{fig:cont-strange}]{
\includegraphics[scale=0.35, clip, trim=0 0.7cm 0 1.4cm]{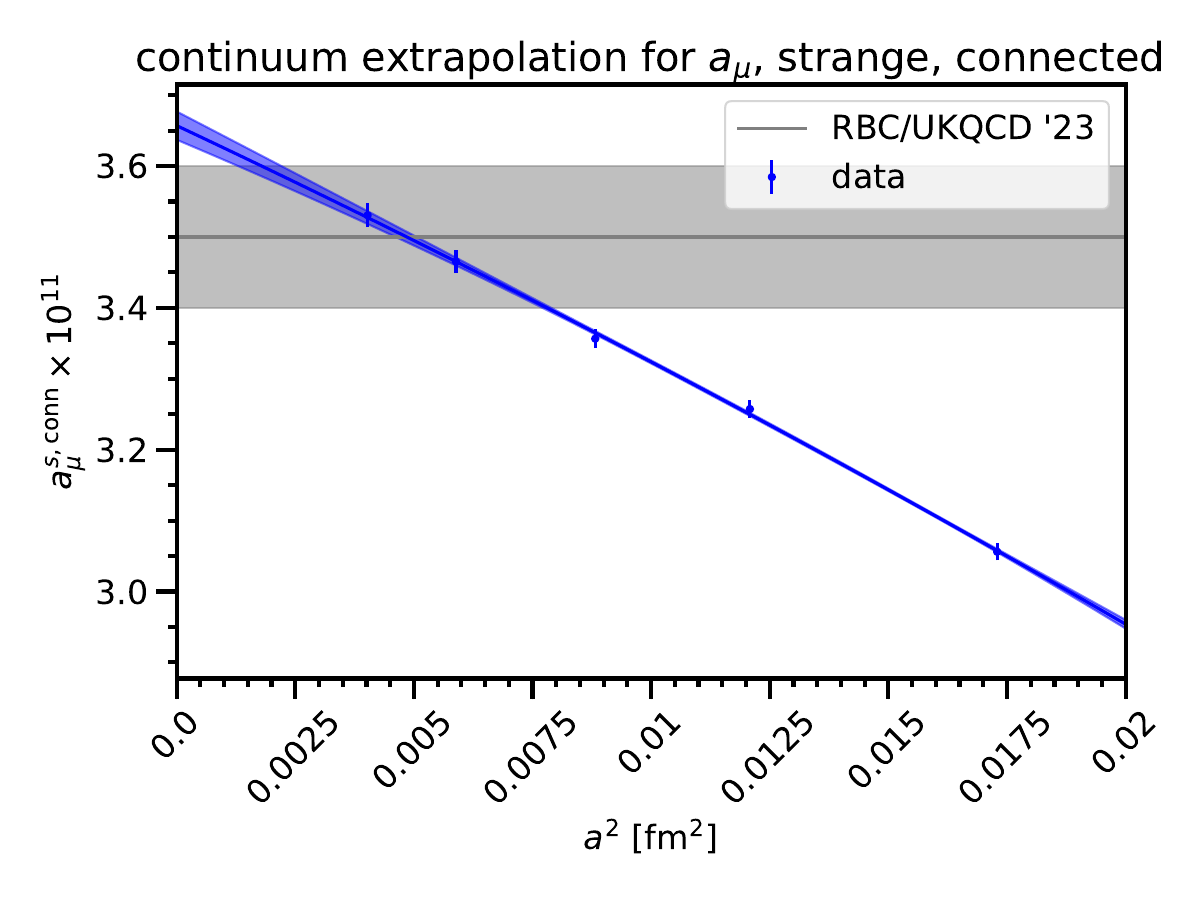}
}
\end{center}
\vspace*{-0.75cm}
\caption{Left: The integrand as a function of $y$ for the connected strange contribution and two different volumes at our coarsest lattice spacing. Right: Continuum extrapolation of the connected strange contribution. We also show the recent RBC/UKQCD value (gray band) \cite{Blum:2023vlm}.\label{fig:strange}}
\end{figure}

In the following, we first consider the connected strange contribution, which has been computed with high precision on several lattice spacings between $0.13~\mathrm{fm}$ and $0.063~\mathrm{fm}$. Figure \ref{fig:finvol-strange} shows the integrand as a function of $|y|$ for two volumes with $L=3~\mathrm{fm}$ and $L=4~\mathrm{fm}$. The integrated results are $a_\mu^{s,\mathrm{conn}}(L\!=\!3~\mathrm{fm}) = 3.056(6)_\mathrm{stat}(12)_{a} \times 10^{-11}$ and $a_\mu^{s,\mathrm{conn}}(L\!=\!4~\mathrm{fm}) = 3.087(6)_\mathrm{stat}(12)_{a} \times 10^{-11}$, where the second error is introduced by the uncertainty of the lattice spacing. The statistical error is already at the per-mille level. Moreover, it appears that finite volume effects are rather small. Taking into account our results for the finer lattices for $L=3~\mathrm{fm}$, we are able to perform a continuum extrapolation, where we use again the fit ansatz \eqref{eq:fit-S2}. Our result for the continuum value is $a_\mu^{s,\mathrm{conn}} = 3.657(20)_{\mathrm{stat}}(14)_{a}$. The extrapolation is plotted in \fig\ref{fig:cont-strange}. Our result is compatible with the recent RBC/UKQCD result \cite{Blum:2023vlm} within $1.5\sigma$. Notice that our analysis does not yet take into account systematic errors like finite volume effects.

\begin{figure}
\begin{center}
\subfigure[connected, light\label{fig:finvol-light-conn}]{
\includegraphics[scale=0.35, clip, trim=0 0.7cm 0 1.3cm]{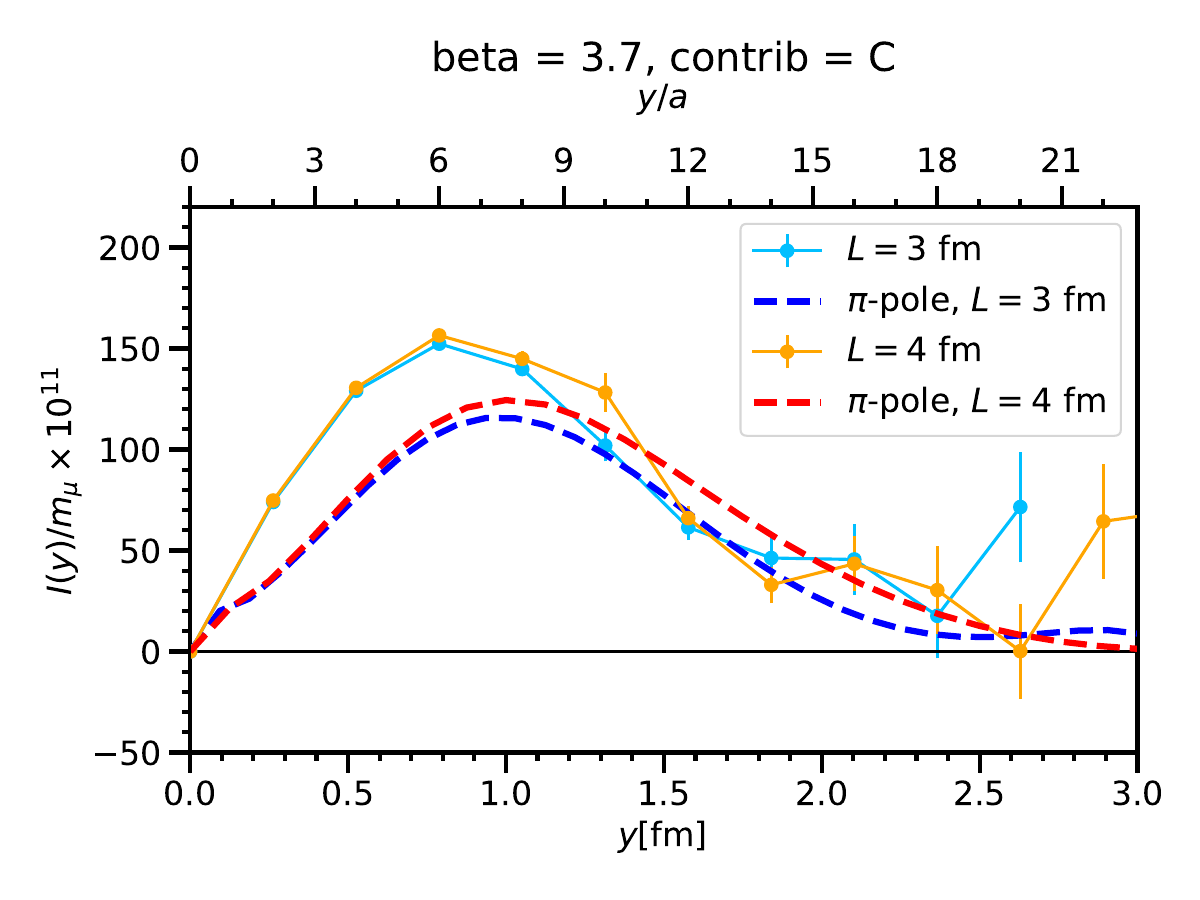}
}
\subfigure[connected + disconnected, light\label{fig:finvol-light-cd}]{
\includegraphics[scale=0.35, clip, trim=0 0.7cm 0 1.3cm]{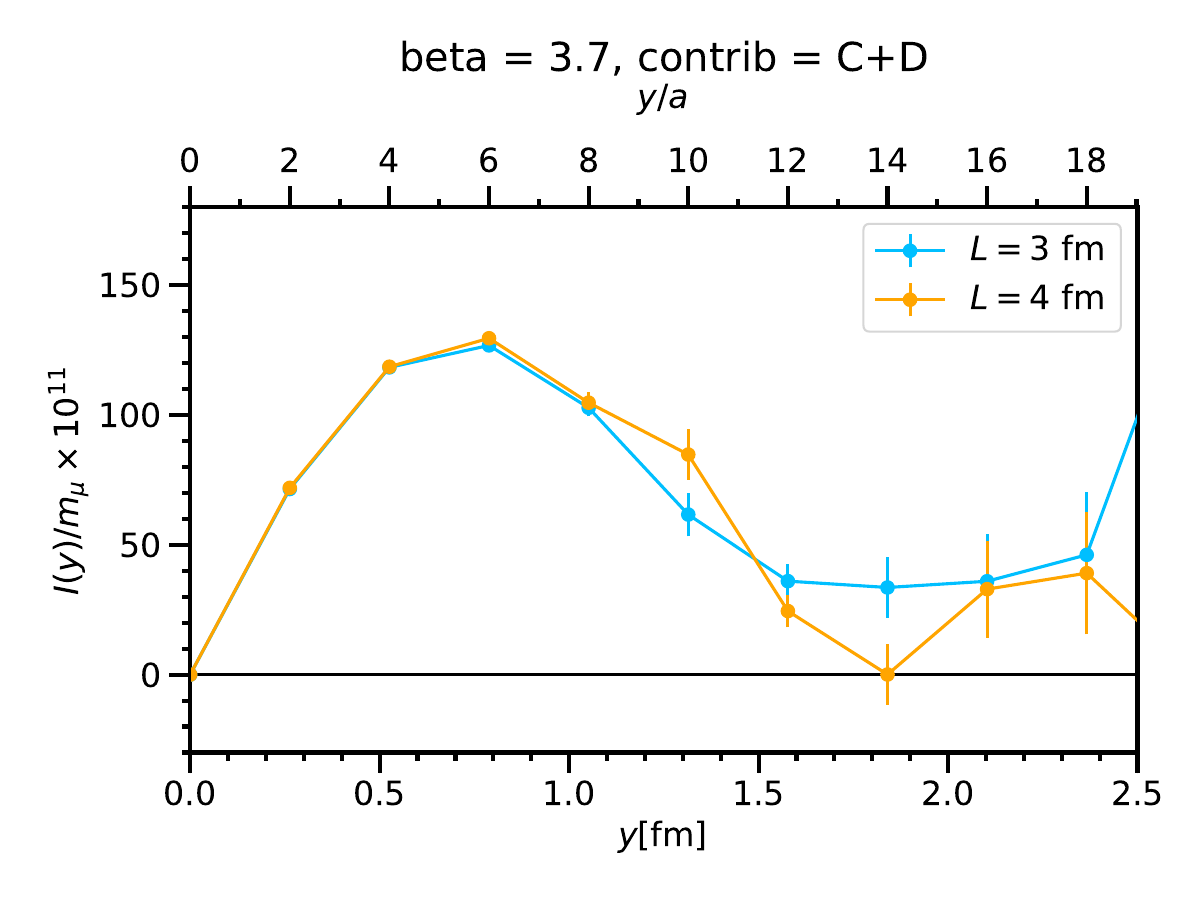}
}
\end{center}
\vspace*{-0.75cm}
\caption{Comparison of the integrand of the connected light contribution for two physical volumes.\label{fig:finvol-light}}
\end{figure}

We now present first results for the light-quark contribution for the coarsest lattices, where we want to focus on the data quality. The integrands are plotted as a function of $|y|$ in \fig\ref{fig:finvol-light} for the connected contribution (a), as well as the sum of connected and disconnected contributions (b). For both we obtain a clear signal with reasonable error. The sum of the two contributions is shown \fig\ref{fig:finvol-light-cd}, where we also show the corresponding result for the $4~\mathrm{fm}$ lattice. Here we can conclude that finite volume effects are moderate. The same applies for the pure connected contribution, which is plotted in \fig\ref{fig:finvol-light-conn}. Since finite volume effects represent one of the most important sources of systematic errors in our study, it is crucial to determine their size. From the dispersive approach we know that the dominant long-range contribution is given by the pion pole \cite{Blum:2023vlm}. Hence, we are able to estimate the finite volume effects by decomposing the four-current correlator \eqref{eq:4c-corr} in terms of pion transition form factors (TFFs):
\vspace*{-0.3cm}

\begin{align}
\widetilde{\Pi}^{(\mathrm{A}, \pi,\mathrm{TFF})}_{\mu\nu\sigma\lambda} (x,y,z) 
&= 
\int_{uv} D_\pi(u-v) \left( M_{\mu\nu}(u,x,y) M_{\sigma\lambda}(v,z,0) + M_{\mu\lambda}(u,x,0) M_{\nu\sigma}(v,y,z) \right) \,,
\nonumber \\
M_{\mu\nu}(u,x,y) 
&:= 
i \epsilon_{\mu\nu\rho\sigma} \partial_{\rho}^{(x)} \partial_{\sigma}^{(y)} \int_{qk} F(q^2,k^2) e^{iq(x-u)} e^{ik(y-u)} \,,
\end{align}
where $D_\pi(u-v)$ is the pion (Klein-Gordon) propagator and $F(q^2,k^2)$ the pion TFF, which has been determined on the same lattices as in the present study \cite{Gerardin:2023naa}. Notice that the pure pion-pole contribution at the integrand level for large $|y|$ and the connected contribution on the lattice are related to each other. Using isospin symmetry one can derive $ I^{\mathrm{conn},\ell}(y) = \frac{34}{9} I^{\pi,\mathrm{latt}}(y)$ \cite{Gerardin:2017ryf}.

The result for the connected contribution reconstructed from the pion-pole calculation is also shown in \fig\ref{fig:finvol-light-conn} for both box sizes. For large distances, they are consistent with the corresponding lattice data. Moreover, the pion pole result clearly predicts a small but non-negligible finite volume effect.

\section{Conclusion}

We considered the hadronic light-by-light contribution to $a_\mu$ on the lattice using staggered fermions and adapted the Mainz method accordingly. The lepton loop result has been successfully reproduced. We have shown some first results on the connected strange-quark contribution. Here we found that finite size effects are small. Moreover, we performed a preliminary continuum extrapolation, the result is compatible with the recent RBC/UKQCD value. Furthermore, we considered the light-quark contribution on our coarsest lattices focusing on finite volume effects. We observe a good data quality, finite volume effects appear to be moderate. The study of finite volume effects is supplemented by considering the pion-pole contribution, which we calculate numerically using our knowledge of the TFF computed on the same set of ensembles. The results are consistent with the lattice data. Finite volume effects are observed to be small but non-negligible.

Our study of the HLbL contribution is still ongoing. It remains to perform the continuum extrapolation for all flavor combinations including light, strange and charm quarks.

\section*{Acknowledgments}

We thank all the members of the Budapest-Marseille-Wuppertal collaboration for helpful discussions and the access to the gauge ensembles used in this work. 
This publication received funding from the Excellence Initiative of Aix-Marseille University - A*Midex, a French ``Investissements d'Avenir" programme, AMX-18-ACE-005 and from the French National Research Agency under the contract ANR-20-CE31-0016. 
The computations were performed on Irene at TGCC. 
We thank GENCI (grants A0080511504, A0100511504 and A0120511504) for awarding us computer time on these machines. 
Centre de Calcul Intensif d'Aix-Marseille (CCIAM) is acknowledged for granting access to its high performance computing resources.

\bibliographystyle{plainnt}
\bibliography{biblio}

\begin{thebibliography}{10}

\bibitem{Muong-2:2006rrc}
G.~W. Bennett et~al.
\newblock {\em Phys. Rev. D}, 73:072003, 2006, {{[arXiv:hep-ex/0602035]}}.

\bibitem{Muong-2:2023cdq}
D.~P. Aguillard et~al.
\newblock 8 2023, {{[arXiv:2308.06230[hep-ex]]}}.

\bibitem{Aoyama:2020ynm}
T.~Aoyama et~al.
\newblock {\em Phys. Rept.}, 887:1--166, 2020, {{[arXiv:2006.04822[hep-ph]]}}.

\bibitem{Borsanyi:2020mff}
Sz. Borsanyi et~al.
\newblock {\em Nature}, 593(7857):51--55, 2021,
  {{[arXiv:2002.12347[hep-lat]]}}.

\bibitem{Blum:2023vlm}
Thomas Blum et~al.
\newblock 4 2023, {{[arXiv:2304.04423[hep-lat]]}}.

\bibitem{Chao:2021tvp}
En-Hung Chao et~al.
\newblock {\em Eur. Phys. J. C}, 81(7):651, 2021,
  {{[arXiv:2104.02632[hep-lat]]}}.

\bibitem{Chao:2020kwq}
En-Hung Chao et~al.
\newblock {\em Eur. Phys. J. C}, 80(9):869, 2020,
  {{[arXiv:2006.16224[hep-lat]]}}.

\bibitem{Asmussen:2022oql}
Nils Asmussen et~al.
\newblock {\em JHEP}, 04:040, 2023, {{[arXiv:2210.12263[hep-lat]]}}.

\bibitem{Lepage:2011vr}
G.~Peter Lepage.
\newblock 11 2011, {{[arXiv:1111.2955[hep-lat]]}}.

\bibitem{Gerardin:2023naa}
Antoine G\'erardin et~al.
\newblock 5 2023, {{[arXiv:2305.04570[hep-lat]]}}.

\bibitem{Gerardin:2017ryf}
Antoine G\'erardin et~al.
\newblock {\em Phys. Rev. D}, 98(7):074501, 2018,
  {{[arXiv:1712.00421[hep-lat]]}}.

\end{thebibliography}

\end{document}